\documentclass[article]{aa}
\usepackage{graphicx}
\usepackage[varg]{txfonts}
\usepackage{amsmath}
\usepackage{hyperref}
\usepackage{aas_macros}
\usepackage{float}

\bibpunct{(}{)}{;}{a}{}{,} 

\begin{document}
\title{A new fitting-function to describe the time evolution of a galaxy's gravitational potential}

\author{Hans J.T. Buist 
\and Amina Helmi}

\institute{Kapteyn Astronomical Institute, University of Groningen, P.O. Box 800, 9700 AV Groningen, The Netherlands\\ \email{buist@astro.rug.nl}}

\date{Received 15 November 2013 /
Accepted 19 January 2014
}

\abstract{We present a new simple functional form to model the evolution of a spherical mass distribution in a cosmological context. Two parameters control the growth of the system and this is modelled using a redshift dependent exponential for the scale mass and scale radius. In this new model, systems form inside out and the mass of a given shell can be made to never decrease, as generally expected. This feature makes it more suitable for studying the smooth growth of galactic potentials or cosmological halos than other parametrizations often used in the literature. This is further confirmed through a comparison to the growth of dark matter halos in the Aquarius simulations.}

\maketitle

\section{Introduction}

The cosmological model predicts significant evolution in the mass
content of galaxies and of their dark matter halos through cosmic
time. This evolution may be directly measurable using 
stellar streams, as these are typically sensitive probes of the
gravitational potential in which they are embedded \citep{Eyre2009,Gomez2010}. Clearly, to properly model this 
dynamical evolution it is critical to use a physically motivated representation of the 
time-dependency of the host's gravitational potential.

In the literature the time evolution of a galactic potential is often parametrized
through the evolution of its characteristic parameters, such as total
mass and scale radius.  In the case of dark matter halos, the most
often used parameters are its virial mass $M_\text{vir}$ and concentration $c$. The virial mass 
is defined as the mass enclosed within the virial radius $r_\text{vir}$, and at this radius
the density of the halo $\rho(r_{\rm vir}) = \Delta \times \rho_{\rm crit}$, where $\rho_{\rm crit}$  is the critical density of the
Universe, and the exact value of $\Delta$  depends on cosmology \citep{NFW1996,NFW1997}.  The
evolution of $M_\text{vir}$ and  $c$ has been thoroughly studied in cosmological numerical
 simulations \citep{Bullock2001,
  Wechsler2002,Zhao2003a,Zhao2003b,Zhao2009,Tasitsiomi2004,BoylanKolchin2010}. It has been found that the 
evolution of the virial mass is well fitted with an exponential in
redshift (but see also \citealt{Tasitsiomi2004} and \citealt{BoylanKolchin2010} for more complex
functions), while the concentration
appears to depend linearly on the expansion factor
(\citealt{Bullock2001,Wechsler2002}, but see also \citealt{Zhao2003a,Zhao2003b,Zhao2009}). These relations work well in a
statistical sense for an ensemble of halos, but do not always
guarantee that the evolution of an individual system is well represented
and physical as we discuss below. Furthermore, it has recently been
pointed out that part of the evolution in mass, especially at late
times, is driven by the definition of virial mass (through its
connection to the background cosmology) rather than to a true increase
in the mass bound to the system \citep{Diemer2013}. 

In this paper, we revisit the most widely used model for the time evolution of 
dark matter halos (Sect.~\ref{sec:w02}) and show that for certain choices of the characteristic parameters, mass growth does not proceed inside out in this model. In Sect.~\ref{sec:new-model} we develop a new prescription for the time evolution of a general spherical mass distribution that does have that property. 
In Sect.~\ref{sec:sims} we compare this model to the growth of dark halos in cosmological simulations
and we summarize in Sect.~\ref{sec:concl}.

\section{Wechsler's model of the evolution of NFW halos}
\label{sec:w02}
The most commonly used approach to model the evolution of dark matter halos is that of \citet{Wechsler2002}. Cosmological simulations show that halos follow characteristic density profiles known as NFW after the seminal work of \citet*{NFW1996}. 
Generally, a two-parameter mass-profile may be expressed as
\begin{equation}
  M(r, t) = M_s\, f(r / r_s) \ ,
\label{eq:massprofile}
\end{equation}
with $r_s(t)$ the scale radius, $M_s(t)$ the mass contained within the scale radius, and $f(x)$ the functional form of the mass profile, with $f(1) = 1$. The relative mass growth rate $\dot{M}/M$ is
\begin{equation}
  \frac{\dot{M}}{M}(r,t) = \frac{\dot{M}_s}{M_s} - \frac{\dot{r}_s}{r_s} \kappa(r/r_s) \ ,
\label{eq:evolutionequation}
\end{equation}
where $\kappa(x)$ is the logarithmic slope of the mass profile
\begin{equation}
  \kappa(x) = \frac{d\log f}{d\log x} \ .
\end{equation}
Typically $\kappa(x)$ is a positive monotonically decreasing function of radius. For the NFW profile, $f(x)$ is given by
\begin{equation}
  f(x) = \frac{A(x)}{A(1)};\ \ \ \ \ \ \  A(x) = \log(1+x) - \frac{x}{1+x} \ .
\end{equation}
The relation between $M_s$ and virial mass $M_\text{vir}(t)$ for an NFW profile is 
\begin{equation}
  M_s = \frac{M_\text{vir}}{f(c)} \ ,
\end{equation}
where the concentration $c \equiv r_\text{vir} / r_s$. As discussed in the Introduction, the virial radius is defined as the radius at which the spherically averaged density reaches a certain threshold value $\rho_c$. For a given choice of $\rho_c$, the virial radius and the virial mass are related through
\begin{equation}
  M_\text{vir} = \frac{4}{3}\pi \rho_\text{c} r_\text{vir}^3 \ .
  \label{eq:virialmass}
\end{equation}
Therefore, only two parameters from the set \{$M_s$, $r_s$, $M_\text{vir}$, $c$\} are needed to fully specify the profile. According to \cite{Wechsler2002} the virial mass evolves as 
\begin{equation}
  M_\text{vir}(z) = M_O \exp{\left[-2 a_c (z - z_O)\right]} \ ,
\end{equation}
with $M_O = M_\text{vir}(z_O)$ and $z_O$ the epoch where the halo is "observed". The formation epoch $a_c$ is arbitrarily defined to be the expansion factor at which $d\log M / d \log a = 2$. \cite{Wechsler2002} found the concentration to evolve on average as
\begin{equation}
  c(a) = 4.1 \frac{a}{a_c}.
\end{equation} 
Halos with more quiescent histories are best described by these relations, while violent mergers can lead to significant
departures from these smooth functions. 

\begin{figure*}[t]
\centering
 \includegraphics[width=1\textwidth]{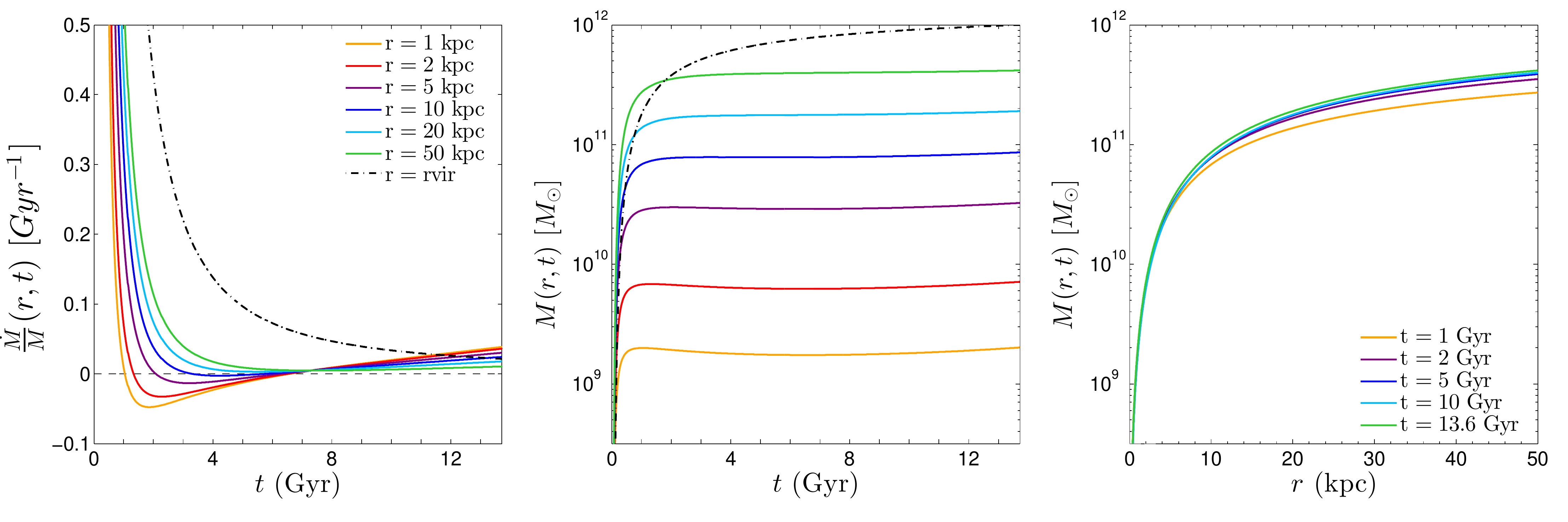}
\caption{Evolution in time of a halo with virial mass $M_\text{vir} = 10^{12} M_\odot$ and formation epoch $a_c=0.15$ (for a cosmology with $h = 0.7$, $\Omega_m = 0.28$, and $\Omega_\Lambda = 0.72$) using the Wechsler model. The left panel shows that the mass growth rate reaches a minimum for all shells between $t=1$ and $t=7$ Gyr, and is negative for the innermost shells which implies they decrease in mass. The central panel shows the mass history for all shells. The shells show a temporary stall in their growth, or even a decrease in mass, between $t=1$ and $t=7$ Gyr, before resuming growth again at later times. The right panel shows the mass profile at different epochs.}
\label{fig:wechslermodel}
\end{figure*}

The Wechsler model can give realistic trajectories in $M_\text{vir}(t)$ and $c(t)$ for many individual halos. However, Fig.~\ref{fig:wechslermodel} shows a problem case. In this figure we have plotted the behaviour predicted for a halo of $10^{12}$ M$_\odot$ and $a_c = 0.15$. The left panel shows the mass growth rate $\dot{M}/{M}$ as a function of time for different shells in the halo. We note that the mass growth rate of a given shell decreases with time, as expected, but that for inner shells it becomes negative, indicating that the mass in the shell has decreased. At later times, the mass growth for the inner shells seems to increase and even exceeds the growth rate at larger radii. 

This behaviour is unexpected in the $\Lambda$CDM cosmology, as halos tend to form inside out \citep{Helmi2003, Wang2011}, implying that the inner shells collapse first and should not grow further at late times by smooth accretion; usually a redistribution of mass is only expected during major mergers. Let us now look in more detail at what causes this odd unphysical behaviour.

From Eq.~(\ref{eq:evolutionequation}), the condition that leads to a violation of $\dot{M}/M > 0$ at a given radius is 
\begin{equation}
  \frac{\dot{M}}{M} = \frac{\dot{M}_s}{M_s} - \kappa(x) \frac{\dot{r}_s}{r_s} < 0\ ,
\end{equation}
which we can express in terms of $M_\text{vir}(t)$ and $c(t)$ using that 
\begin{equation}
 \frac{\dot{M_s}}{M_s} =  \frac{\dot{M}_{\rm vir}}{M_{\rm vir}} - \kappa(c) \frac{\dot{c}}{c}
\end{equation}
and 
\begin{equation}
\frac{\dot{r}_s}{r_s} = \frac{\dot{r}_\text{vir}}{r_\text{vir}} - \frac{\dot{c}}{c}
 = \frac{\dot{M}_{\rm vir}}{3 M_{\rm vir}} - \frac{\dot{\rho}_c}{3 \rho_c} - \frac{\dot{c}}{c} \ .
\end{equation}
This leads to
\begin{equation}
  \frac{\dot{M}}{M} = \frac{\dot{M}_{\rm vir}}{M_{\rm vir}}\left(1 - \frac{\kappa(x)}{3}\right) + 
\frac{\dot{c}}{c} (\kappa(x) - \kappa(c)) + \frac{\kappa(x)}{3} \frac{\dot{\rho}_c}{\rho_c} \ .
\label{eq:timemvir}
\end{equation}
Since for an NFW it can be shown that $\kappa(x) \le 2$, for the Wechsler model, the first term is
\begin{equation}
 \frac{\dot{M}_{\rm vir}}{M_{\rm vir}}\Big(1 - \frac{\kappa(x)}{3}\Big) = -2 a_c \dot{z} \Big(1 - \frac{\kappa(x)}{3}\Big) > 0 \ .
\end{equation}
On the other hand, since $\kappa(x)$ is a monotonically decreasing function, the second term is
\begin{equation}
 \frac{\dot{c}}{c}\Big(\kappa(x) - \kappa(c)\Big) = -\frac{\dot{z}}{1 + z}  \Big(\kappa(x) - \kappa(c)\Big) > 0\ \ \ (r < r_\text{rvir}) \ .
\end{equation}
Finally, the last term in Eq.~(\ref{eq:timemvir}) is always negative as $\dot{\rho_c}/\rho_c < 0$. This implies that there may be times and radii for which the combination of the various terms is negative, leading to a decrease in the mass contained within a given shell. The above analysis shows that the Wechsler model has this behaviour because of pseudo-evolution: the virial mass is defined with respect to the cosmological background density and this background density evolves in time.

\section{Alternative model}
\label{sec:new-model}

\begin{figure*}
\centering
 \includegraphics[width=1\textwidth]{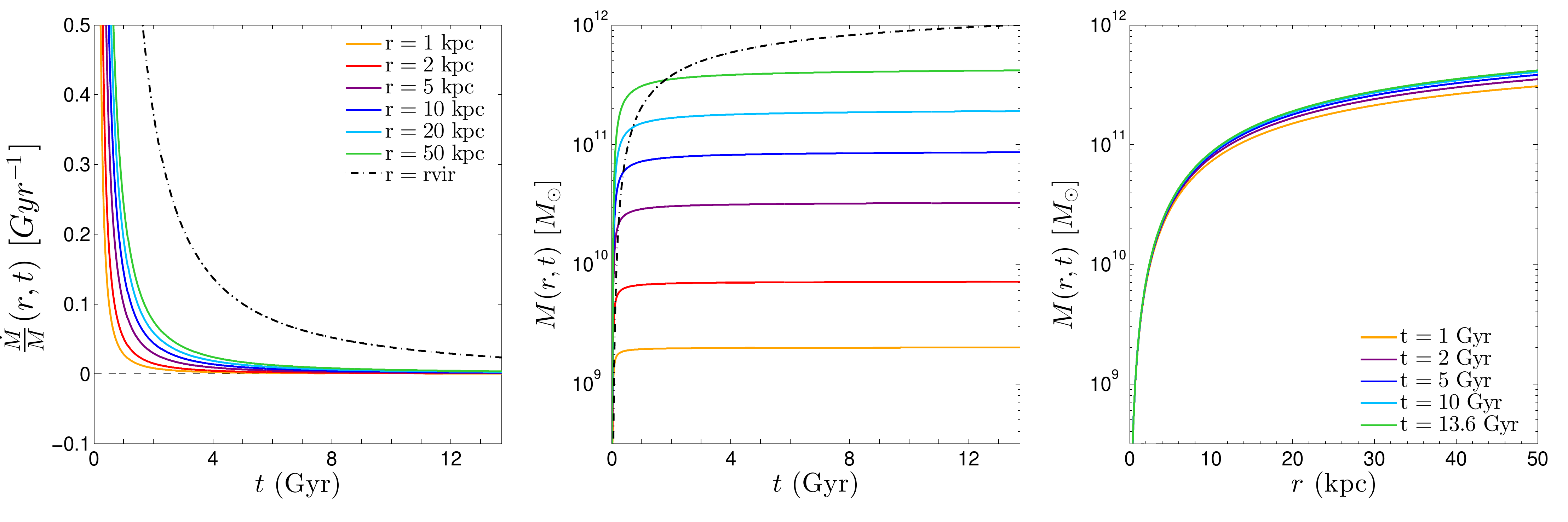}
\caption{Evolution in time of the same final halo of Fig.~\ref{fig:wechslermodel}, but using our new model with $a_g = 0.04$ and $\gamma=2$. Left panel: the relative growth rate does not reach a minimum, nor does it reach negative values. This implies that the shells grow at all times, but as the central panel shows, the mass growth history for the shells flattens out towards the end. The behaviour at the virial radius follows closely (but not exactly) the Wechsler model, as shown by the dash-dotted curve. Right panel: Mass profile at different epochs.}
\label{fig:ncfwmodel}
\end{figure*}

We have explored a different way of modelling the evolution of a mass profile, by directly focusing on the growth of the scale radius $r_s(t)$ and the scale mass $M_s(t)$. We require our model to have non-negative mass growth at all times and at every radius
\begin{equation}
  \frac{dM}{dt} \ge 0, \, \forall \ r, t \ .
\end{equation}
In addition, we require that the mass grows inside out, which means that the relative mass growth rate should increase with radius 
\begin{equation}
  \frac{\partial}{\partial r}\left(\frac{\dot{M}}{M}\right) \ge 0, \, \forall \ r, t \ .
\end{equation}

The logarithmic slope $\kappa(x)$ in Eq.~(\ref{eq:evolutionequation}) is essential in the determination of whether these conditions are satisfied. Because $\kappa(x)$ is a monotonically decreasing positive function, we can set a lower limit on the mass growth
\begin{equation}
  \frac{\dot{M}}{M} \ge \frac{\dot{M_s}}{M_s} - \kappa_{\rm max} \frac{\dot{r_s}}{r_s}\ , 
\end{equation}
where $\kappa_\text{max}$ is the maximum value of $\kappa(x)$. For NFW and Hernquist profiles $\kappa_\text{max} = 2$, for Jaffe profiles $\kappa_\text{max} = 1$, while for the Plummer and isochrone potentials $\kappa_\text{max} = 3$. The condition $\dot{M}/M = 0$ would lead
to a solution of the form
\begin{equation}
  M_s(t) = M_{s,O} \left(\frac{r_s(t)}{r_{s,O}}\right)^{\kappa_\text{max}}\ ,
\end{equation}
with $r_{s,O}$ and $M_{s,O}$ the scale mass and scale radius at some epoch $z_O$. Motivated by this we propose that in general, a solution of Eq.~(\ref{eq:evolutionequation}) should be of the form
\begin{equation}
  M_s(t) = M_{s,O} \left(\frac{r_s(t)}{r_{s,O}}\right)^{\gamma}\ ,
  \label{eq:powerlawrelation}
\end{equation}
where $\gamma > 0$ is a constant. This functional form is also supported by the work of \cite{Zhao2003a,Zhao2003b}, who used N-body cosmological simulations to show that a strong correlation between $r_s$ and $M_s$ exists. This correlation may be modelled as a power law with exponent $3\alpha$ (so $\gamma = 3\alpha$) for NFW profiles. The growth equation is then given by
\begin{equation}
  \frac{\dot{M}}{M} = \frac{\dot{M_s}}{M_s} \left(1 - \frac{\kappa(r/r_s)}{\gamma} \right) \ .
\end{equation}
Assuming that $\dot{M}_s \ge 0$, the mass growth rate is positive at every radius for $\gamma \ge \kappa_\text{max}$, while for $\gamma < \kappa_\text{max}$ the mass will decrease at radii where $\gamma < \kappa(r/r_s)$. The critical radius $r_{\rm crit}$ occurs where $\gamma = \kappa(r_{\rm crit} /r_s)$. Since at early times $r_s \ll 1$, all radii are beyond the critical radius and the mass increases everywhere. As the scale radius grows, eventually some radii move inside the critical radius. When this occurs, the mass at those radii would actually decrease according to this model, but this can be avoided with a proper choice of $\gamma$. 

This model also ensures that the mass grows inside out since 
\begin{equation}
  \frac{\partial}{\partial r}\left(\frac{\dot{M}}{M}\right) = - \frac{\dot{M_s}}{M_s} \frac{1}{\gamma} \frac{\partial \kappa(r/r_s)}{\partial r} \ge 0\ , \ \ (\gamma \ge 0)
\end{equation}
and the gradient of the logarithmic slope of $\kappa(x)$ is negative.

For the evolution of the parameters $M_s(t)$ and $r_s(t)$ we propose an exponential behaviour similar to that used by \citet{Wechsler2002}, i.e.
\begin{equation}
  M_s(z) = M_{s,O} \exp{\left[-2 a_g (z - z_O)\right]} \ ,
\label{eq:mass_z}
\end{equation}
with $a_g$ the growth parameter of the halo. Using Eq.~(\ref{eq:powerlawrelation}), the evolution of $r_s$ is given by
\begin{equation}
  r_s(z) = r_{s,O} \exp{\left[-2 \frac{a_g}{\gamma} (z - z_O)\right]} \ .
\label{eq:rs_z}
\end{equation}
This model has two free parameters (after one has fixed $z_O$), namely $a_g$ which determines the growth of $M_s$, and $\gamma$, 
which is related to the growth rate of the scale radius through $a_g/\gamma$. Fig. \ref{fig:ncfwmodel} shows the evolution of 
a $10^{12} M_\odot$ halo that has the same final mass and scale radius as that in Fig.~\ref{fig:wechslermodel}, taking $a_g = 0.04 $ and $\gamma = 2$ to produce a similar growth history. 
The evolution of the halo is as desired: the mass growth rate (left panel) is always positive and slows down as time goes by, and this slowing down is more pronounced for the innermost shells. The halo thus forms inside out as expected (central panel), and there is no mass exchange/decrease between neighbouring shells. In this figure we have also plotted the behaviour of $M_\text{vir}(t)$, determined using Eqs.~(\ref{eq:massprofile}) and (\ref{eq:virialmass}). We note with satisfaction that the evolution of the virial mass for our model is very similar to that of \citet{Wechsler2002}. 

\begin{figure}
\centering
 \includegraphics[width=0.5\textwidth]{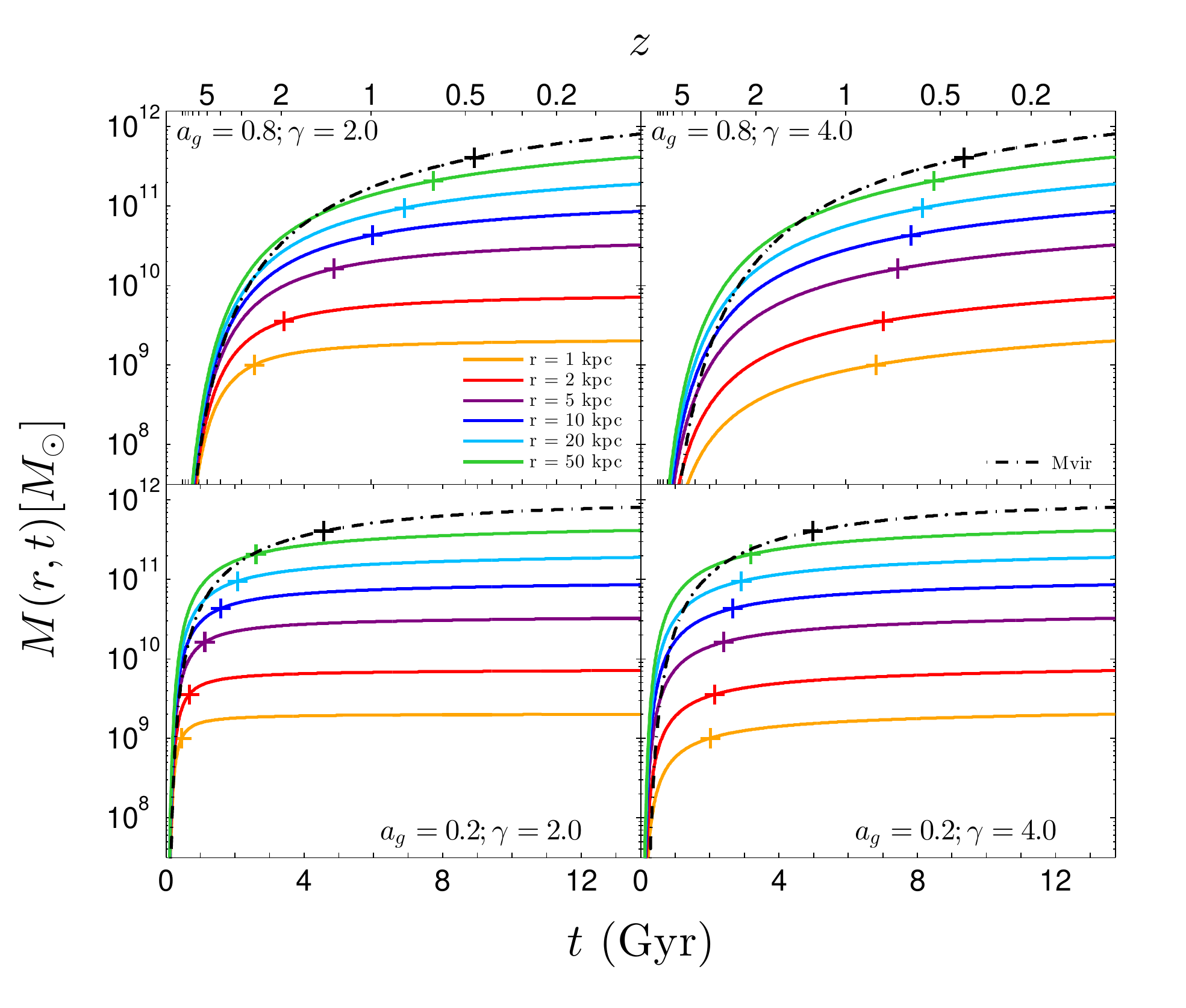}
\caption{Mass growth history for various radial shells, normalized to the final value as in Fig. \ref{fig:ncfwmodel}, for different values of $a_g$ and $\gamma$. The formation epoch $a_g$ changes the growth of the individual shells and the overall growth pattern of the halo. This can also be seen in the half mass time of the shells indicated with a plus on every curve. The parameter $\gamma$ mostly influences how individual shells grow. This is seen as we change $\gamma$ from $2$ to $4$ (left vs right panels): only a small effect is seen in the overall growth of the halo, but the inner shells grow on much shorter timescales than the outer shells.}
\label{fig:ncfwmodelexamples}
\end{figure}

In Fig. \ref{fig:ncfwmodelexamples} we show the effect of changing the slope $\gamma$ and the formation epoch $a_g$. For example, for smaller $a_g$ (bottom panels) the mass growth occurs earlier, and the growth rate today is lower. The parameter $\gamma$ does not alter the overall evolution of the halo very much, but changes the (relative) growth pattern for the individual radial shells. For example, for $\gamma = 2$  (left panels) the inner shells grow much faster than the outer shells, while for $\gamma = 4$ (right panels) the difference in mass growth between the different shells is much smaller. 

\section{Comparison with simulations}
\label{sec:sims}

\begin{figure*}
\label{fig:aquarius}
\centering
 \includegraphics[width=1\textwidth]{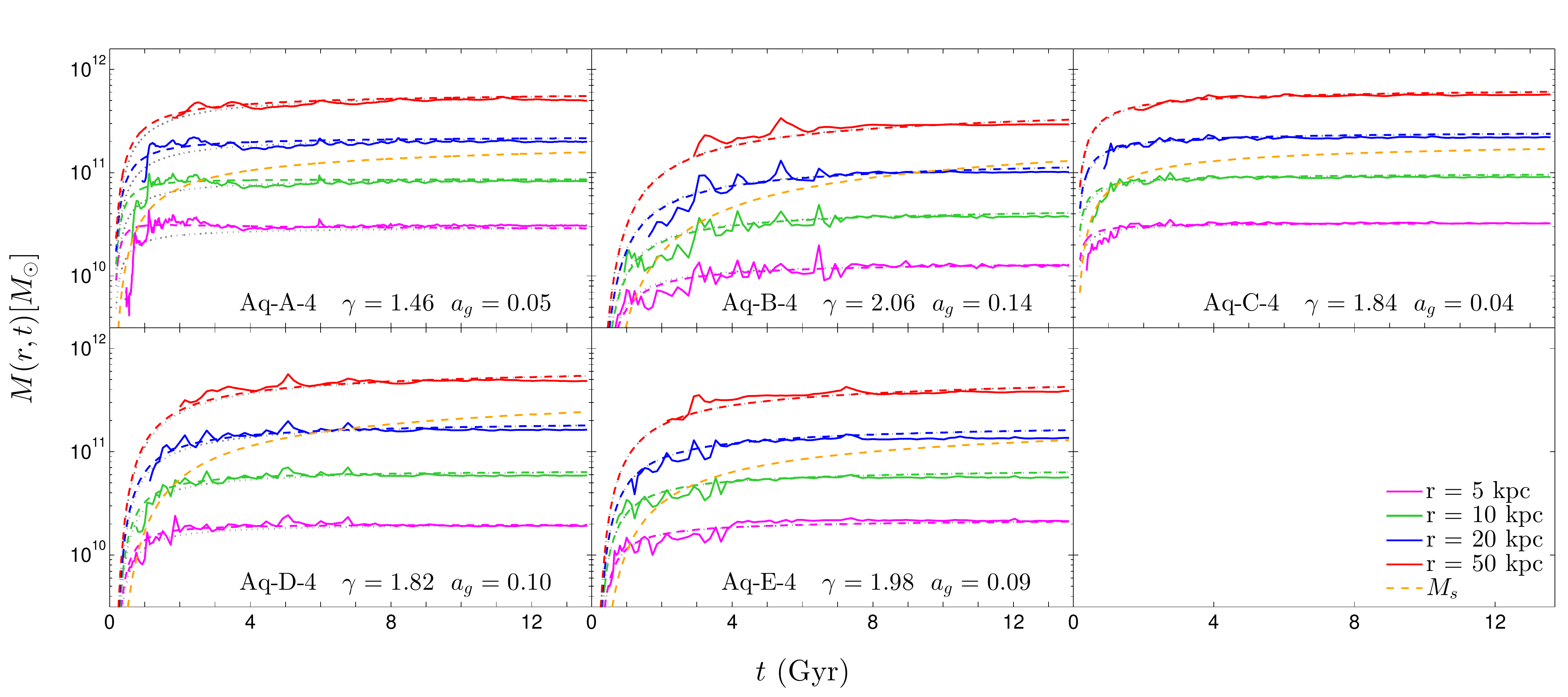}
\caption{Mass growth history of the Milky Way-like Aquarius halos at different radii compared to the fit obtained using our model for $r_s$ and $M_s$ (colour dashed curves) and fitted for $z < 6$ to avoid the epoch of significant  mergers. The dashed orange line indicates the evolution of the scale mass $M_s$. Setting $\gamma=2$ also gives a reasonable fit as indicated by the grey dotted curve for each halo.}
\end{figure*}

To further justify our model, we have studied how well it describes the behaviour of the Milky Way-like dark matter halos from the Aquarius project simulations \citep{Springel2008}. This is a suite of six cosmological dark matter N-body simulations that were run at a variety of resolutions. We used an intermediate resolution level, in which the Aquarius Milky Way-like (or main) halos have $\sim 5 \times 10^6$ particles, and we focused on the behaviour of halos Aq-A-4 to Aq-E-4 (the sixth halo, Aq-F, experiences a major merger at low redshift and so is not considered in our analysis). For each main halo, we computed the spherically averaged density profile at each output redshift and fitted the NFW functional form with parameters $M_s$ and $r_s$. We only used snapshots with $z < 6$ to avoid the epoch of major merger activity. In this way we determine the behaviour of $M_s$ and $r_s$ with redshift for each halo, to which we then fit our model to determine $a_g$ and $\gamma / a_g$ in Eqs.~(\ref{eq:mass_z}) and (\ref{eq:rs_z}). 

The mass growth history of the halos and the results of our fitting procedure are shown in Fig. \ref{fig:aquarius}. The values of $a_g$ for halos Aq-A and Aq-C are low (0.05 and 0.04, respectively), implying an early formation, while Aq-B, Aq-D, and Aq-E having formed later, have larger $a_g$ (0.14, 0.10, and 0.09, respectively). This is consistent with the results of  \cite{BoylanKolchin2010}, who studied the mass accretion history of Milky Way-mass halos in the Millenium-II cosmological simulations. These authors found that halos D and E follow the median history, while halos A and C form earlier. Halo B was found to catch up with the median growth around $z\approx2$. In view of this, we may argue that a value $a_g \sim 0.1$ roughly corresponds to the median mass accretion history of Milky Way-mass halos. 

The values of the slope $\gamma$ determined in our fits are close to $2$, the $\kappa_\text{max}$ for the NFW potential, but generally tend to be slightly smaller, which would imply a mass decrease for some shells. Figure \ref{fig:aquarius} shows that halo Aq-A depicts this behaviour early in its history. Nonetheless, we find that for all halos, setting $\gamma = 2$ also produces a reasonable fit to the evolution history in $M_s$ and $r_s$, as can be seen from the grey dotted line in this figure.

In order to relate our results more closely to the \citet{Wechsler2002} model, we determined the $a_c$ values for halos Aq-A to Aq-E, and found these to be in the range $0.1$ to $0.2$, therefore appearing at the lower end of the distribution given by Wechsler et al. in their Fig. 8. This could be due to the more quiescent merger history of the Aq-halos.

\section{Summary and conclusions}
\label{sec:concl}

We have studied the evolution of a spherical mass distribution in a
cosmological context. Since we expect galaxies (and their halos) to
grow inside out, inner shells should form earlier than outer ones, and
in the smooth accretion regime, their mass should not decrease with
time. Motivated by violations of these conditions found in some models
often used in the literature for certain regions of parameter space,
we have presented an alternative way
of modelling the mass evolution of a spherical potential. In our
model, we let the scale radius and the mass at the scale radius grow
exponentially with redshift, but relate them in such a way that the condition
of no mass decrease at any radius can always be satisfied.  The setup is
quite general, and can be applied to any spherical density profile. The model has two parameters that can be chosen to obtain a variety of
growth histories, providing more control over the growth rate of shells at
different radii.

In comparison to
previous work, our model does not differ greatly from that presented
by \cite{Zhao2003a,Zhao2003b,Zhao2009} who also use a power-law
relation between scale mass and scale radius to describe the growth of
dark matter halos in cosmological simulations, except that we have found a general condition
under which a system following a given density profile will grow inside out.  
\cite{Tasitsiomi2004} found that some accretion histories were
better fitted using a power law of the scale factor $a$, instead of an exponential in redshift
$z$, and proposed a mass accretion history that combines both the
power-law relation and the exponential behaviour. \cite{BoylanKolchin2010} proposed a slight modification to
this model, which gives even better fits to the MS-II
simulations. Their addition of an extra parameter allows the halos to
grow earlier and to reach a much lower growth rate at later times. 

However, given that the comparison to the growth of dark matter halos in the Aquarius
simulations shows that our model works quite well, we believe it is worthwhile to keep its simplicity, particularly in applications that model the behaviour of streams in time-dependent Galactic potentials \citep{Bullock2005,Gomez2010,Gomez2010b}.  \\

We are grateful to Volker Springel, Simon White, and Carlos
Frenk for generously allowing us to use the Aquarius
simulations. Carlos Vera-Ciro is acknowledged for his help in the
analysis of these simulations. H.J.T.B. and A.H. gratefully
acknowledge financial support from ERC-Starting Grant
GALACTICA-240271.

\bibliographystyle{aa} 
\bibliography{article.bib} 




\end{document}